\documentclass[prl,english,aps,prb,a4paper,twocolumn,floatfix,superscriptaddress]{revtex4}
\usepackage[T1]{fontenc}
\usepackage{color}
\usepackage[latin1]{inputenc}
\usepackage{graphicx}
\usepackage{bm}
\usepackage{epsfig}
\usepackage{rotating}
\usepackage{float}

\usepackage[normalem]{ulem}

\usepackage{dcolumn}
\usepackage{bm}

\definecolor{fistro}{rgb}{0.0, 0.5, 0.0}
\definecolor{orange}{rgb}{1, 0.5, 0.3}

\begin{document}

\title{Defects in vertically vibrated monolayers of cylinders}

\author{Miguel Gonz\'alez-Pinto}
\affiliation{Departamento de F\'{\i}sica Te\'orica de la Materia Condensada, Universidad Aut\'onoma de Madrid,
E-28049, Madrid, Spain}  

\author{Johannes Renner}
\affiliation{Theoretische Physik II, Physikalisches Institut, Universit\"at Bayreuth, D-95440 Bayreuth, Germany}    

\author{Daniel de las Heras}
\email{delasheras.daniel@gmail.com}
\affiliation{Theoretische Physik II, Physikalisches Institut, Universit\"at Bayreuth, D-95440 Bayreuth, Germany}  

\author{Yuri Mart\'{\i}nez-Rat\'on}
\email{yuri@math.uc3m.es}
\affiliation{
	Grupo Interdisciplinar de Sistemas Complejos (GISC), Departamento
	de Matem\'aticas, Escuela Polit\'ecnica Superior, Universidad Carlos III de Madrid,
	Avenida de la Universidad 30, E-28911, Legan\'es, Madrid, Spain}

\author{Enrique Velasco}
\email{enrique.velasco@uam.es}
\affiliation{Departamento de F\'{\i}sica Te\'orica de la Materia Condensada,
Instituto de F\'{\i}sica de la Materia Condensada (IFIMAC) and Instituto de Ciencia de 
Materiales Nicol\'as Cabrera,
Universidad Aut\'onoma de Madrid,
E-28049, Madrid, Spain}  

\date{\today}

\begin{abstract}
We analyse liquid-crystalline ordering in vertically vibrated monolayers of cylinders 
confined in a circular cavity. Short cylinders 
form tetratic arrangements with C$_4$ symmetry.
This symmetry, which is incompatible with the geometry of the 
cavity, is restored by the presence of four point defects with total topological charge $+4$.
Equilibrium Monte Carlo simulations predict the same structure.
A new method to measure the elastic properties of the tetratic medium is developed which
exploits the clear similarities between the vibrated dissipative system and the thermal equilibrium 
system. Our observations open up a new avenue to investigate the formation of defects in response to boundary 
conditions, an issue which is very difficult to realize in colloidal or molecular systems.
\end{abstract}


\maketitle

Monolayers of vertically-vibrated grains have been shown to exhibit a surprisingly rich behaviour 
\cite{Narayan,Galanis1,Galanis2,Aranson1,Aranson2}, including pattern formation in steady-state structures and
non-equilibrium phenomena \cite{Narayan2}. Even though experimental control parameters, such as vibration frequency and amplitude, critically affect 
the observed phenomena,
in some regions of parameter space non-equilibrium behaviour is absent or not predominant, and observed patterns resemble those typical 
of interacting 
particle systems in thermal equilibrium. Of particular interest are the patterns exhibited by elongated particles with 
cylindrical shape, which project on a plane approximately as hard rectangles (HR); when their aspect ratio is moderate,
these particles arrange into two-dimensional monolayers with strong tetratic correlations \cite{Narayan,Mueller,us1}. The tetratic phase is a 
liquid-crystalline arrangement with
particles aligned preferentially along two equivalent perpendicular orientations with global C$_4$ symmetry. 
Mean-field density-functional theories for rectangular particles in thermal equilibrium had predicted the existence of this phase \cite{Schlacken,us0}, which
was confirmed experimentally \cite{Chaikin} and by simulation \cite{Donev,us2}. It is remarkable that the same symmetry 
has also been observed in monolayers 
of vibrated granular rods \cite{Narayan,Mueller,us1}. Some theoretical ideas have been advanced to explain general similarities between thermal and granular systems 
\cite{Edwards,Aranson1}, but none of them is based on deep physical roots. 
In the case of systems exhibiting liquid-crystalline order 
only plausible mechanisms that drive locally ordered arrangements of particles can be invoked \cite{us1}.
Also, from the experimental evidence, it is tempting to use arguments based on equilibrium entropy
maximisation and excluded-volume ideas to qualitatively explain the effective driving forces that lead to the 
extended ordered domains observed in the non-equilibrium vibrated granular systems, like in the corresponding 
equilibrium fluid of two-dimensional HR.

\begin{figure*}
\begin{center}
\includegraphics[width=0.90\linewidth,angle=0]{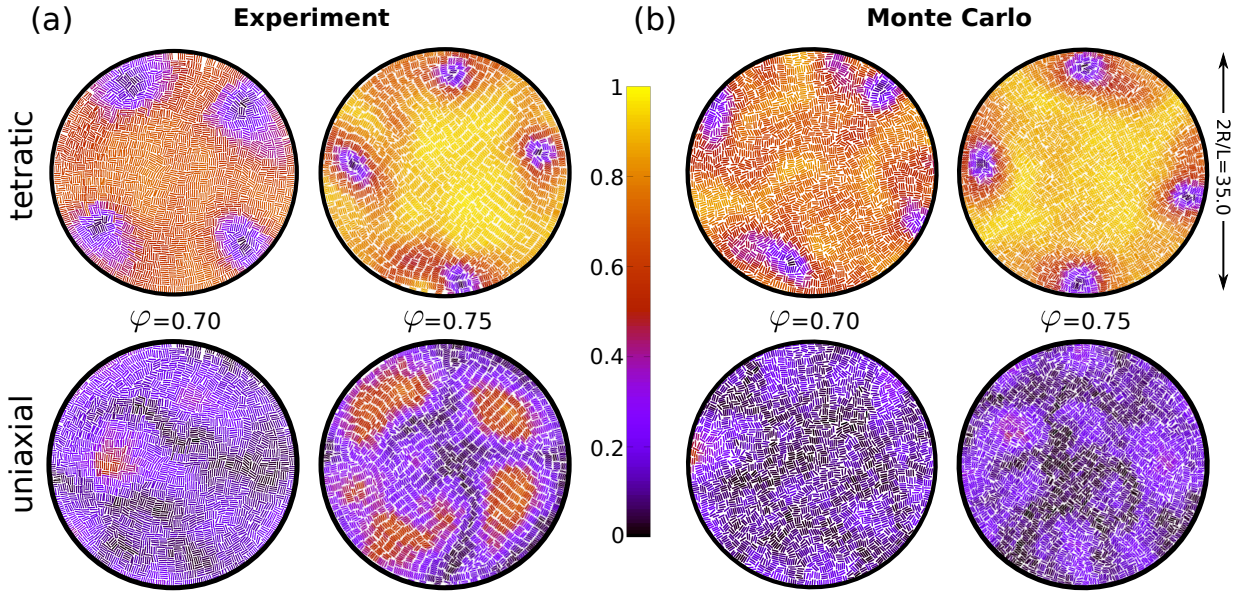}
\caption{\label{fig1} Colour maps of the local tetratic $q_4$ and uniaxial $q_2$
order parameters. The colour of each particle represents the value of the order parameter as specified in
the central colour bar. Data for the cases $\varphi = 0.70$ and $0.75$, $\kappa=4$ and $2R/L=35.0$
from (a) the experiment, and (b) MC simulation.
Particles have been slightly shortened
to improve visualization, and holes in (a) correspond to particles that could not be identified
by the imaging software.}
\end{center}
\end{figure*}

Before a complete theoretical framework for vibrated granular two-dimensional particles is formulated,
more evidence on their similarities with thermal particles is building up. In our previous paper \cite{us1},
we briefly reported on a yet another intriguing behaviour of vibrated granular matter, this time in a monolayer of cylinders 
contained in a quasi-two-dimensional circular cavity. We showed that the tetratic monolayer exhibits four defects symmetrically 
located at the corners of a square. 

In the present paper we analyse this experimental system in detail and show that this behaviour is exactly what is expected in
an equivalent thermal system: our Monte Carlo (MC) simulations of such a system give rise to exactly the same structure \cite{LiBowick}.
Therefore the granular monolayer can respond to geometrical frustration 
much like what one would expect in a thermal system, i.e.
by creating localised defects that interact through the elastic stiffness of the fluid tetratic medium. 
Also, we show that the defected structure is stable in time and occurs in a range of densities. Finally, we propose a method to
calculate the effective interaction between defects in an equilibrium system, and from this the elastic stiffness 
modulus of the liquid-crystal medium. Assuming that the granular system can be treated in the
same manner, the resulting modulus turns out to be of the same order of magnitude in both thermal and granular systems.
Our conclusion is that not only their local order and their response to geometric frustration in terms of defect formation are similar, but also that the elastic properties 
that mediate long-range interactions between the defects are similar.


From the evidence gathered by previous experiments \cite{Narayan,Mueller,us1}, it is clear that
the configurations of cylinders of low length ($L$)-to-width 
($D$) ratio $\kappa=L/D$ change, as
packing fraction $\varphi=\rho LD$
(where $\rho=N/A$ is the effective density, with $N$ the number of cylinders and
$A$ the area) is increased, from orientationally disordered, or {\it isotropic}, fluid arrangements, to {\it tetratic}
fluid configurations. 
Tetratic configurations persist for aspect ratios up to $\kappa\simeq 7$; this limit is also predicted by Monte Carlo (MC) simulations of HR
in thermal equilibrium \cite{Mueller} and supported by density-functional theories \cite{us}. 
At even higher densities smectic fluctuations can
also be seen in the experiment \cite{us1}.


\textbf{Experiment}\\
In the experiment (see Ref. \cite{us1} for more details) 
cylinders made of nonmagnetic steel with
length 4 mm and  width 1 mm ($\kappa=L/D=4$) are placed inside a horizontal circular cavity 
made of aluminium and covered from above by a circular methacrylate lid 
resulting in a free height of 1.8 mm and radius $R=7$ cm ($R/L=17.5$). 
Tne sample is mechanically 
agitated using an electromagnetic shaker which generates a sine-wave vertical motion 
of frequency $\nu=37$ Hz and amplitude $a_0$ with an effective acceleration 
$\Gamma=4\pi^2 a_0\nu/g\simeq 2$, with $g$ the gravity's acceleration. The images were 
taken with a carefully collimated digital camera during the whole duration of the 
experiment (roughly 3 hours).  

Particle identification (position and orientation) 
is done using the ImageJ \cite{ImageJ} software supplemented by our own image processing code. 
Three types of particle arrangements are observed: isotropic, where particles are disordered in both 
orientations and positions; tetratic, where particles show fluid behaviour but are oriented on average 
along two equivalent, perpendicular directions; and smectic, with particles forming fluid layers. 
These configurations are identified by means of order parameters 
$q_n=\left<\cos{n\vartheta}\right>$, with $n=2,4$ (respectively uniaxial and tetratic order parameters), 
and $q_s=\left<e^{i{\bm q}\cdot{\bm r}}\right>$ (smectic order parameter) on each particle,
where $\vartheta$ is the angle of the particle with respect to the local alignment direction $\hat{\bm n}$, 
${\bm r}$ its position and ${\bm q}$ a wavevector compatible with the cylinder length. 
In locally tetratic configurations $q_2<q_4$ and $q_s\sim 0$; 
isotropic and smectic configurations are identified by 
$q_2\sim q_4\sim 0$, $q_s\simeq 0$ and $q_2>q_4$, $q_s>0$, respectively. 
Note that, in our experiments, extended uniaxial nematic configurations are not
formed for any value of aspect ratio and, due to geometric frustration and excitation of vorticity 
smectic domains are limited in size and time \cite{us1}.

\textbf{Simulation}

Two types of simulations have been performed in this work. First, we conducted equilibrium MC simulations on a system of HR that
intends to mimic the experimental granular system, i.e. using the same particle aspect ratio, cavity radius and packing fraction.
Each particle is characterized by its position and orientation vector. $N$ particles are placed in cavity of radius $R/L=17.5$.
The interaction between the particles and the cavity is a hard potential acting on the corners of the particles. That is, the
wall-particle potential is infinity if at least one corner of the particle is outside the cavity and vanishes otherwise. 

Following the ideas of Ref.~\cite{Heras} we initialize the system at very low packing fraction $\varphi=0.1$ for which the equilibrium state
is isotropic. Next we adiabatically increase the number of particles in steps of $\Delta\varphi=0.05$ until the desired packing
fraction is reached. After every increment in the number of particles we run $10^6$ Monte Carlo Sweeps (MCS) to equilibrate
the system. Each MCS is an attempt to sequentially move and rotate every particle in the system. The maximum rotation and displacement
each particle is allowed to perform in one MCS is recalculated in each simulation such that the acceptance probability of the motion is $\sim0.25$.
In order to insert new particles we randomly select one particle in the cavity and create a replica with the same orientation but displaced by $\sim D$ 
along the long access. If the new particle overlap with other particles or with the cavity we reject the insertion and select another random particle.
Once a new particle is accepted we perform a few hundred rotations and translations on it. 
The orientational order parameters and other quantities of interest, such as the positions of the defects, are calculated following
the same procedures as for the experiments.

Second, we performed Brownian simulations on systems of four point particles inside a circular cavity. These
effective particles represent point defects of the real system. The effective particles 
interact with each other via a repulsive pair potential $u(s)=-c\log{(s/a)}$, where $a$ is an irrelevant length-scale, $s$ is the interdefect distance,  
and $c$ is a strength parameter related
to the stiffness coefficient. The use of a logarithmic pair potential can be justified
by invoking the interaction that results from the solution of the Frank elastic theory
\cite{Bowick}. In the elastic theory the constant $a$ is the characteristic 
dimension of the defect core. However at the dynamical level this constant does not play any role 
because forces are not affected by its value. Therefore we used $a=R$ as a 
length-scale. In addition, the effect of the cavity surface is introduced through 
a potential 
$V(r)=\epsilon\exp{[-\lambda(R-r)]}$, 
where $\epsilon$, and $\lambda$ 
are the strength and the inverse decay-length of the wall potential. 
Defects are assumed to be subject to thermal fluctuations from particles in the 
tetratic medium, driven by a thermal energy $k_BT$. These fluctuations are taken care of by solving a Langevin 
equation $m\dot{\bm v}_i={\bm F}_i-\zeta{\bm v}_i+{\bm g}_i$, where $m$ is an effective
mass, ${\bm v}_i$ the velocity of the $i$th defect, ${\bm F}_i$ the force on the defect from 
the total potential $V(r_i)+\displaystyle\sum_{j\ne i}u(r_{ij})$, $\zeta$ a friction
coefficient, and ${\bm g}_i$ a stochastic white noise. Using the cavity radius $R$ and the parameter $c$ as length and energy
scales, respectively, the dimensionless 
discretised equation, in the non-inertial, Brownian regime, becomes
\begin{eqnarray}
	{\bm r}_i^*(t+h)={\bm r}_i^*(t) +\gamma{\bm F}_i^*(t)+\sqrt{2\gamma T^*}{\bm\eta}_i(t),
\end{eqnarray}
where ${\bm r}_i^*(t)$ is the position of the $i$th defect in units of $R$
at time $t$,
${\bm F}_i^*$ is the force on the $i$th defect in units of $c/R$, 
$T^*=k_BT/c$, where $T$ is an effective temperature, $\gamma=hc/R^2\zeta$,  
$h$ the time step, and ${\bm\eta}_i$ a dimensionless Gaussian 
noise of unit variance and zero mean. The free parameters of the model are $T^*$, 
$\lambda^*=\lambda R$, and $\epsilon^*=\epsilon/c$. Since the simulations involve only four
particles they can be extended for several hundred millions time steps to collect
statistically significant information.
---------------------------\\

\begin{figure}
\begin{center}
\includegraphics[width=0.90\linewidth,angle=0]{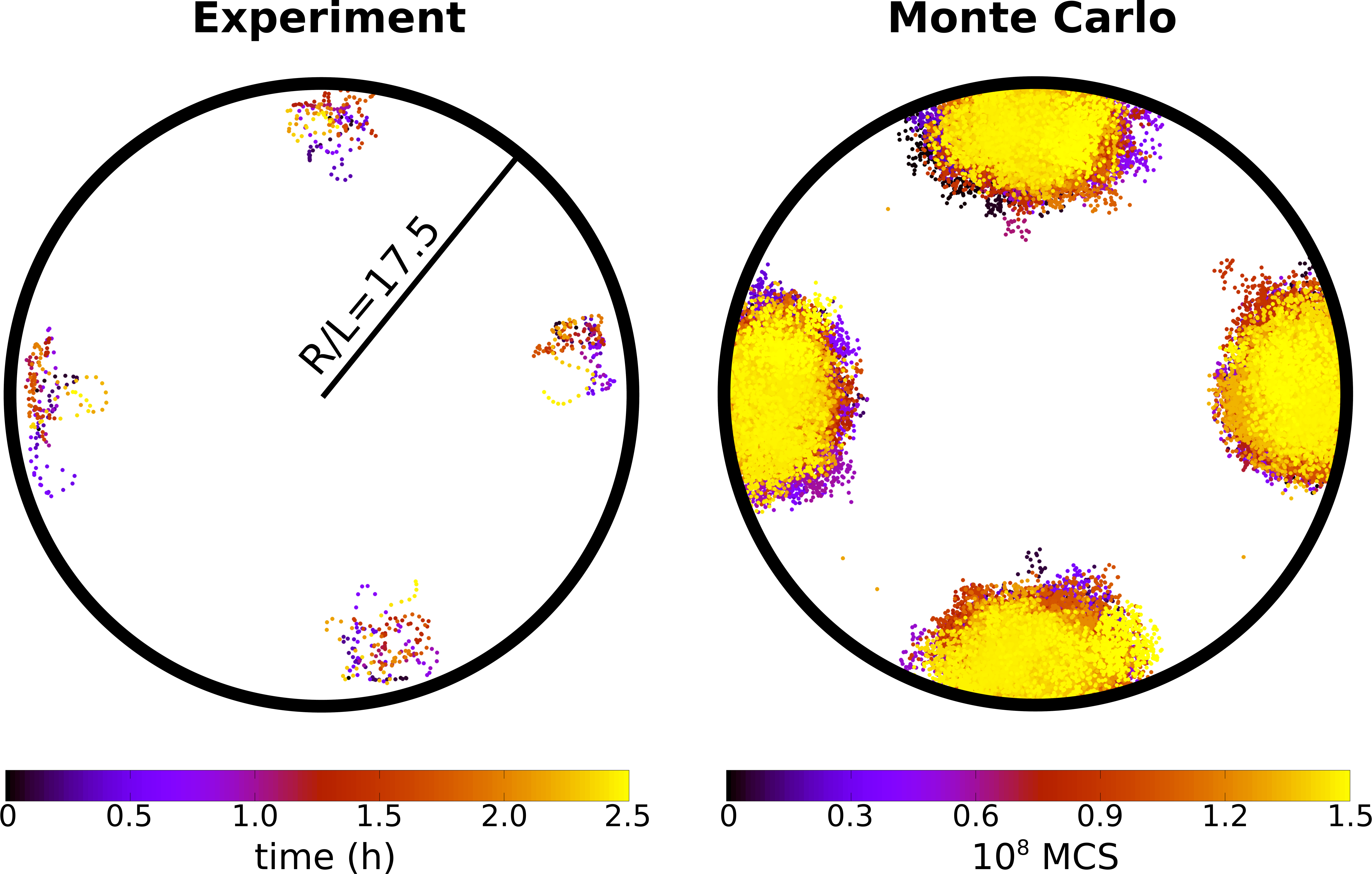}
\caption{\label{fig2} Sampling of defect positions in experiment (left) and Monte Carlo simulation (right).
Colour bars below indicate the experimental time in hours, and the `MC time' in units of
$10^8$ MC sweeps.}
\end{center}
\end{figure}

\begin{figure*}
\begin{center}
\includegraphics[width=0.3\linewidth,angle=0]{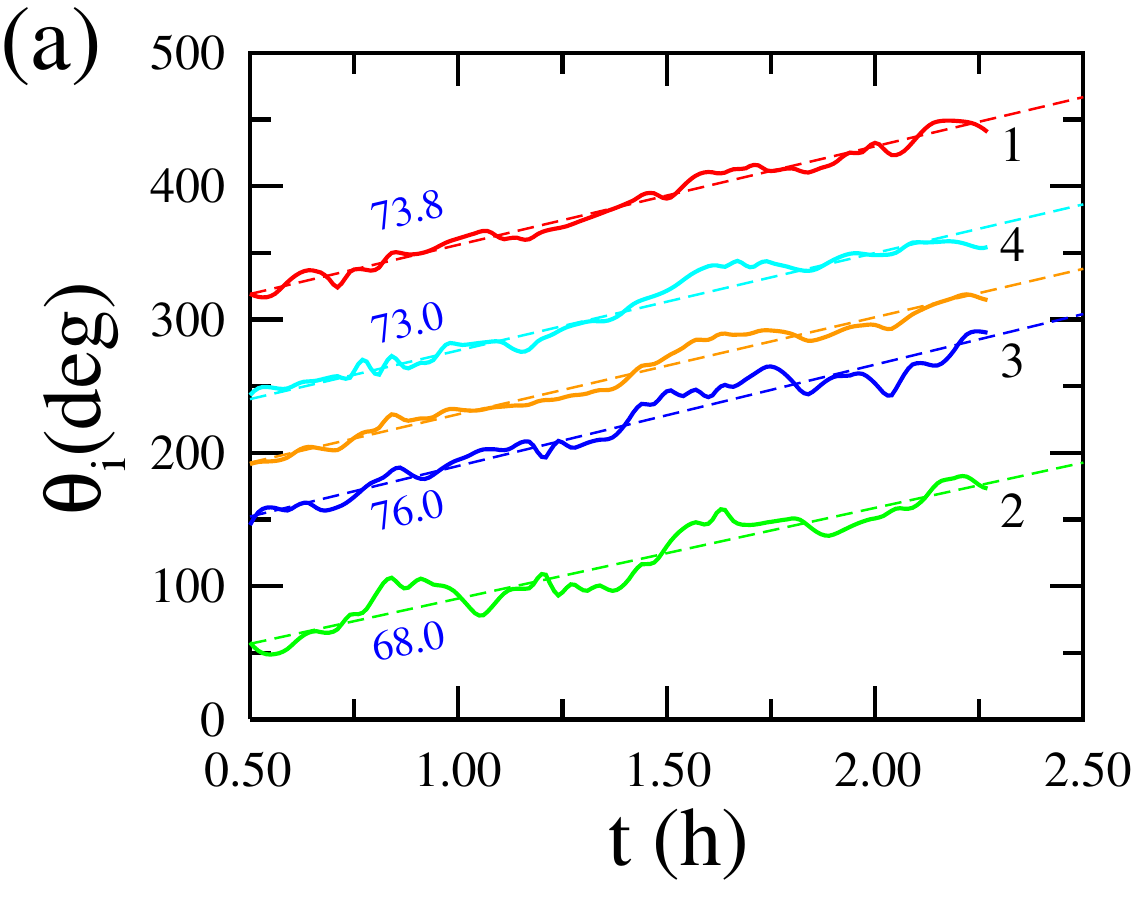}
\includegraphics[width=0.32\linewidth,angle=0]{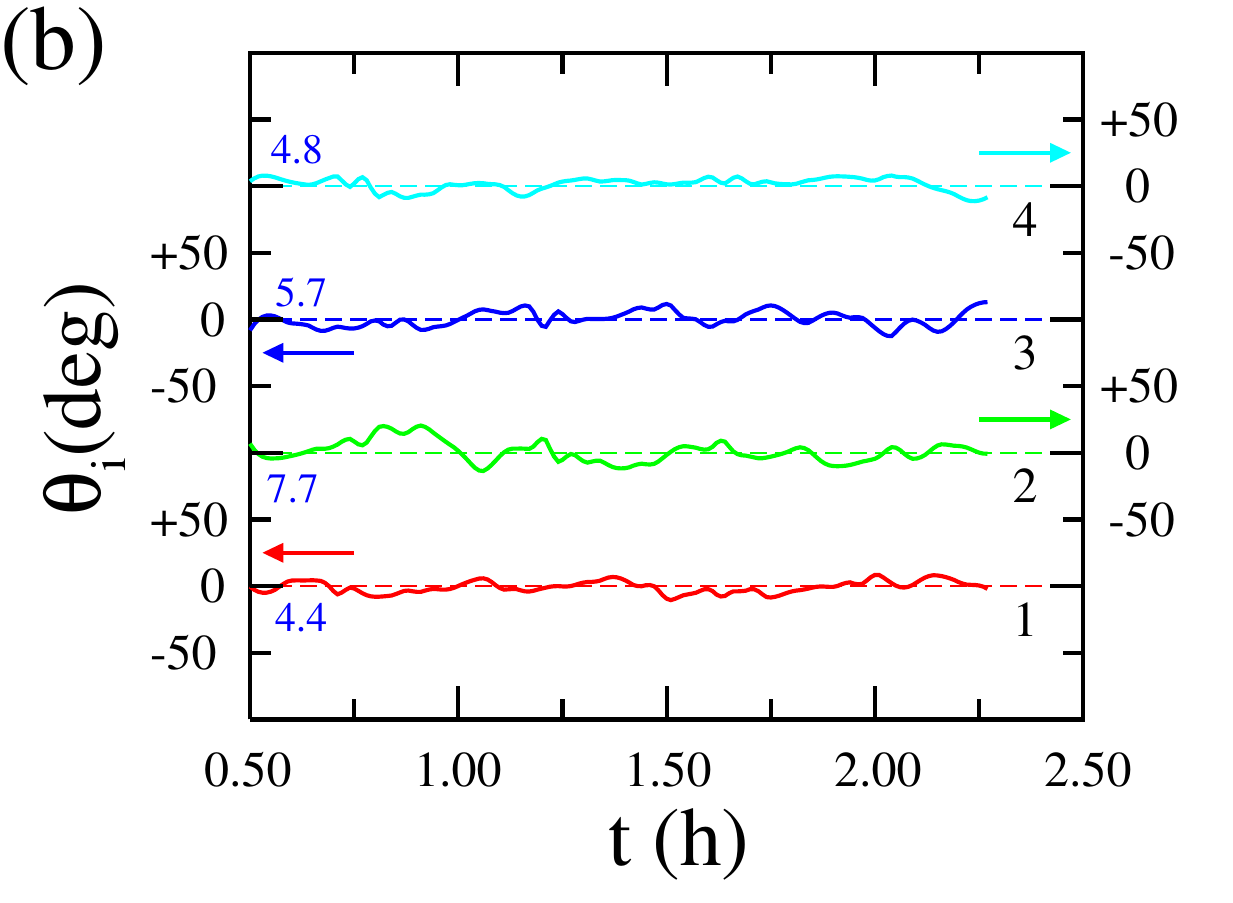}
\includegraphics[width=0.32\linewidth,angle=0]{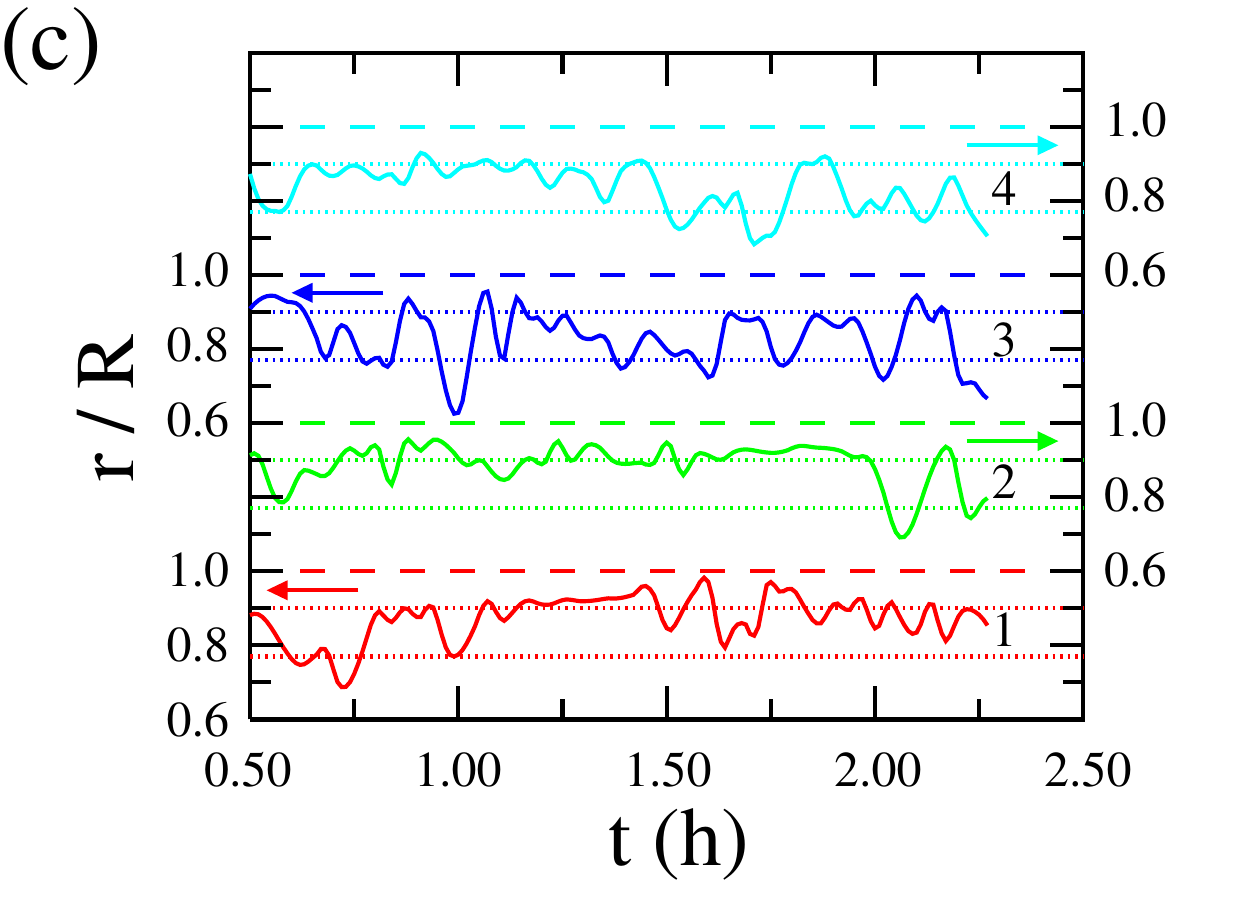}
\includegraphics[width=0.30\linewidth,angle=0]{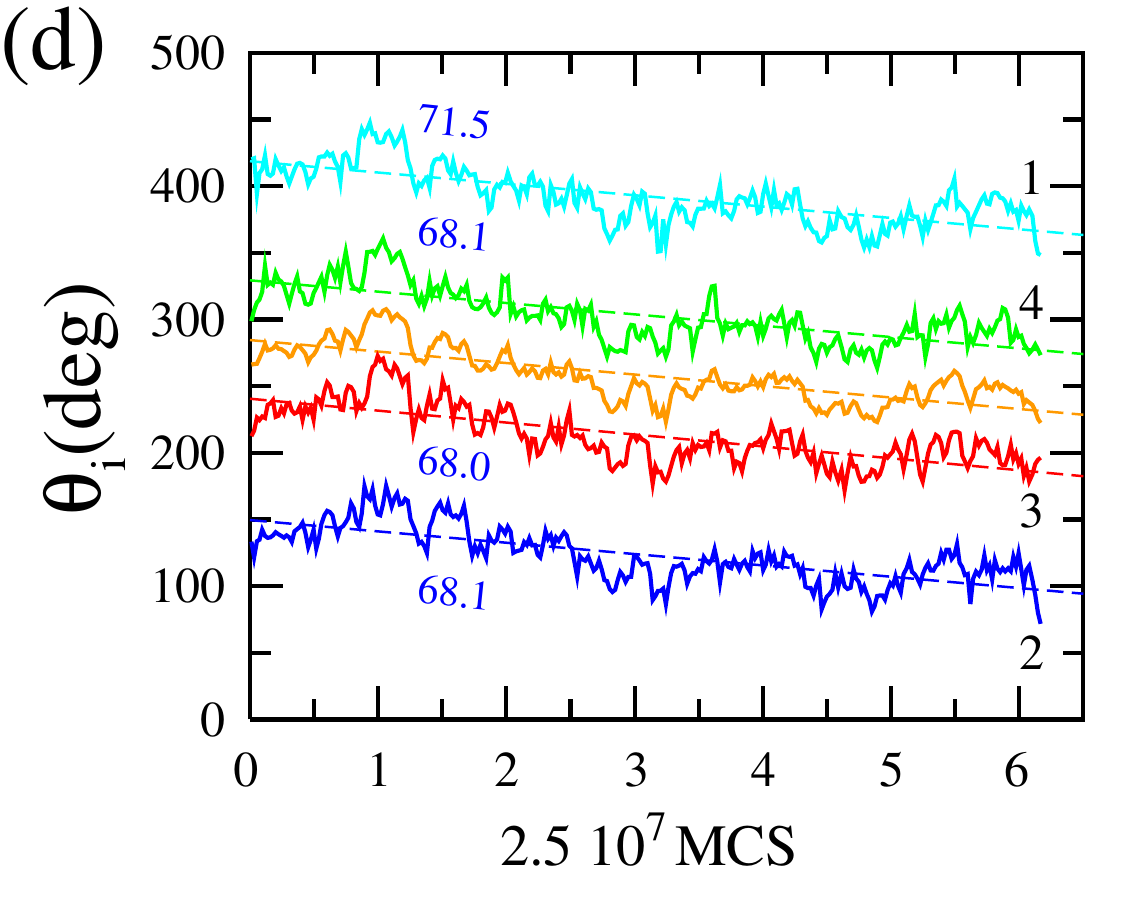}
\includegraphics[width=0.32\linewidth,angle=0]{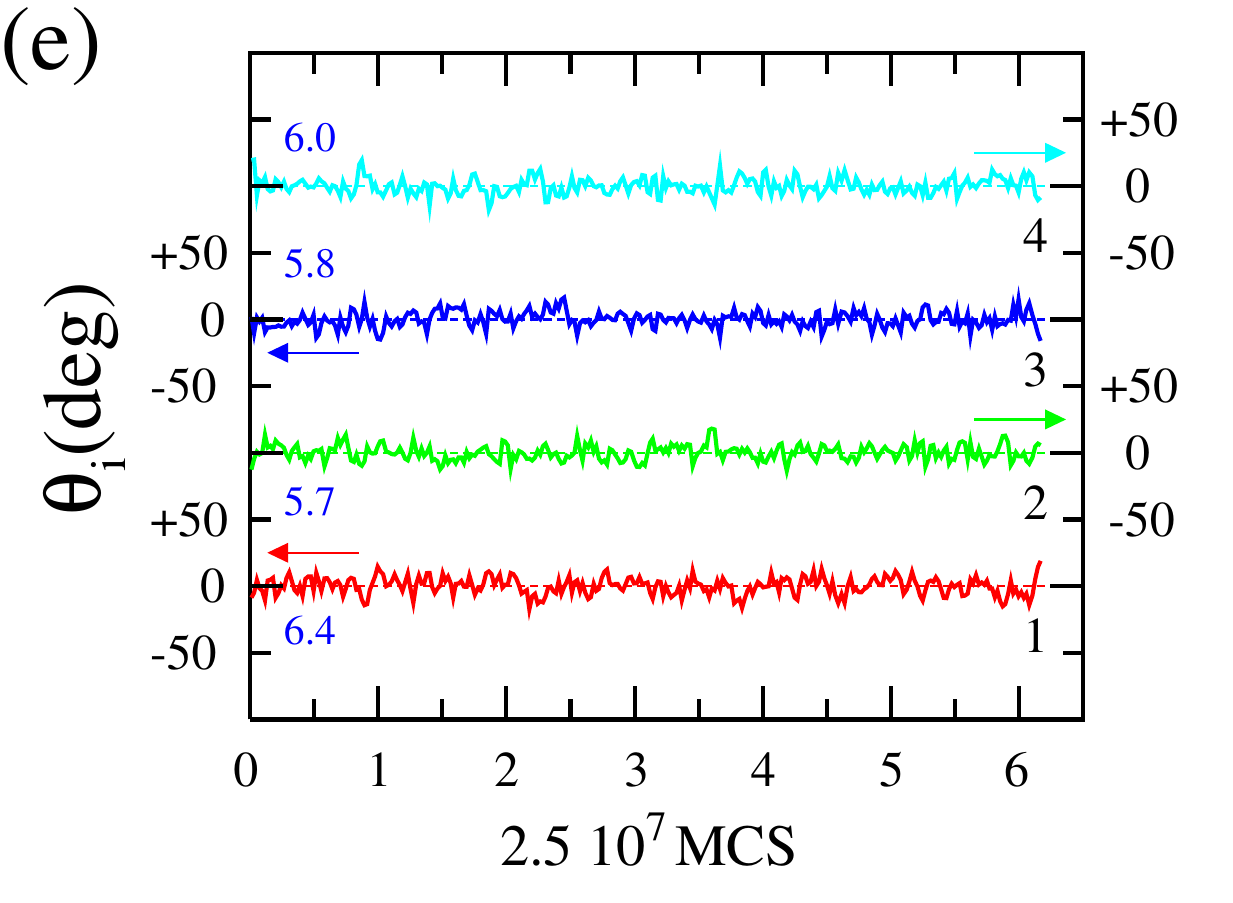}
\includegraphics[width=0.32\linewidth,angle=0]{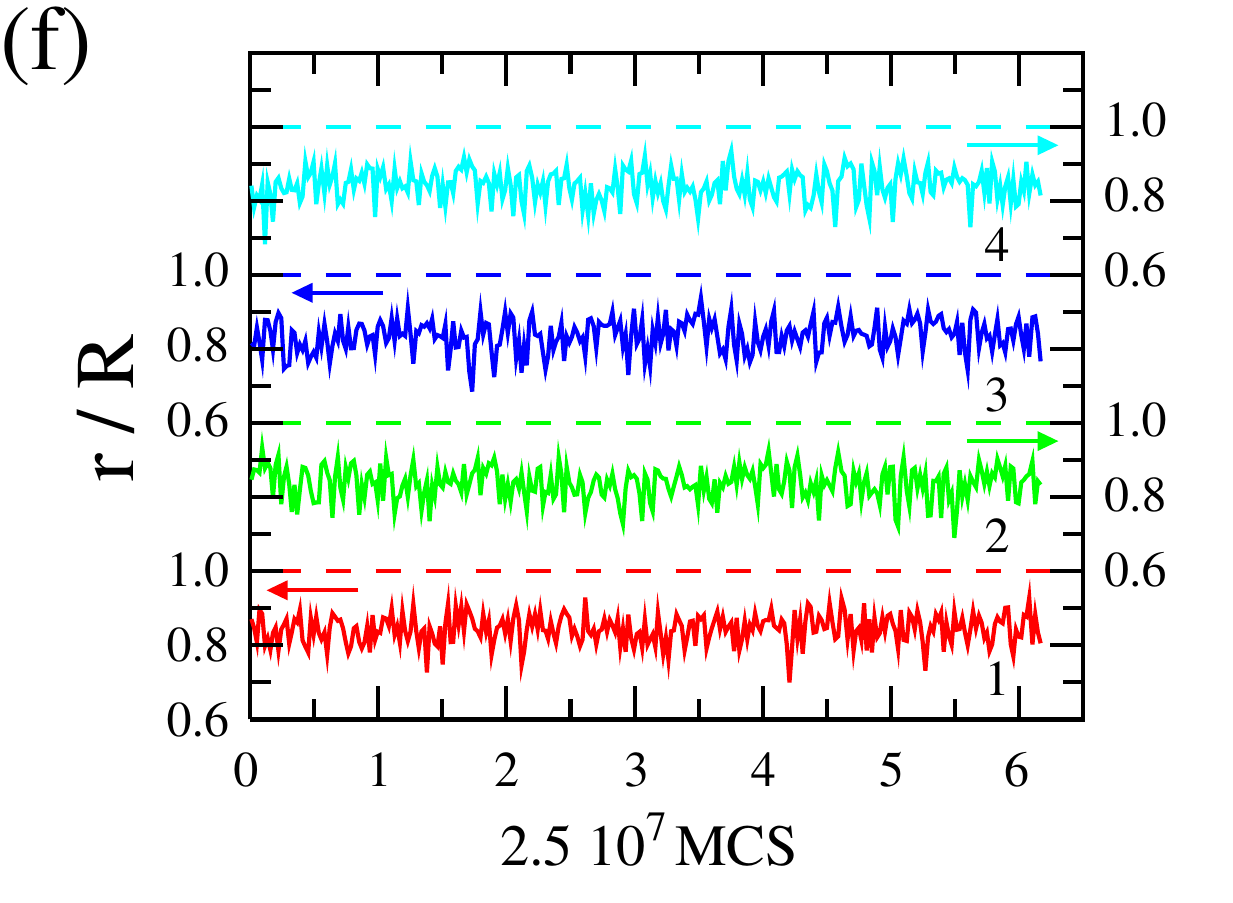}
	\caption{\label{fig3} (a)-(c) Experimental positions of the four defects as a function of time (given in hours). 
Data for each defect are displaced for clarity. (a) Azimuthal angles $\theta_i$ in the lab frame for each defect
(indicated as labels). The average angle is represented by the orange curve.
Dashed lines are linear fits, while labels in blue indicate the values of the slope in degrees per hour. (b) Same as in (a), 
but angles referred to the instantaneous rotating frame (curve in orange in
panel (a)). Labels indicate
variance in each case. (c) Defect radial distances $r/R$ scaled with cavity radius. 
Dashed lines indicate the edge of the cavity in each case, while dotted lines correspond to the two maxima
	of radial distance distribution, Fig. \ref{fig4}(b). Panels (d)-(f) show the same magnitudes as obtained 
	from MC simulations. In (d) labels indicate the absolute
value of the slope in degrees
per MCS, multiplied arbitrarily by a factor $2\times 10^8$ to make it
of the same order of magnitude as in panel (a).} 
\end{center}
\end{figure*}

Fig. \ref{fig1} shows particle configurations from (a) our experiment and (b) Monte Carlo simulation.  In each 
case two values of the packing fraction, $\varphi=0.70$ and $0.75$, are shown. The two orientational order parameters, 
uniaxial $q_2$ and tetratic
$q_4$, have been coded in false colour on each particle, using the same protocols in both experiment and simulation.
To obtain the order parameter fields at some point ${\bm r}$ we average over the particles 
located within a circular region of radius $\xi=4L$ centered at ${\bm r}$ as 
discussed in Ref. \cite{us1}.

The structure in the cavity is similar in the two cases, experiment and thermal model. Also, tetratic configurations are visible at the two densities,
as indicated by the high values of the tetratic order parameter (see central colour bar) and the low values of the uniaxial order 
parameter. In these `fluid' configurations, which are very stable, particles are globally oriented along two perpendicular
directions, with approximately half of the particles pointing in each direction. This balance holds locally, except in
the high-density experimental configuration shown, which exhibits clear patches where the $q_2$ order parameter
is enhanced; these correspond to domains where particles point along a common direction and organise into `smectic' layers. 
Smectic domains fluctuate in time in a sea of particles with tetratic 
order. When present, as in the configuration shown, they adopt a four-fold symmetry.
In the $q_4$ maps four regions containing particles with a $q_4$--depleted
neighbourhood are also clearly visible. These regions are always present in this range of densities.
They can be assimilated to point defects 
of topological charge ${\cal Q}_i=+1$ ($i=1,\dots,4$). The presence of these four defects 
restores the globally tetratic symmetry, which is broken by the circular geometry of the cavity. The total charge $+4$ satisfies 
the  constraint imposed by the Gauss-Bonnet theorem for a medium with $C_4$ symmetry. 
Note that these numbers fulfill the equation $\sum_i{\cal Q}_i=p\chi$, with $p=4$ the $p-$fold symmetry of the tetratic 
phase, while $\chi=1$ is the Euler characteristic of the disk \cite{Bowick}. The same four-defect structure is visible
in the MC simulations. Indeed the thermal
system reacts to confinement in the same way as the granular system, i.e. by creating four point defects symmetrically
located next to the wall. In this case, however, smectic domains are never seen; instead, the whole cavity is filled
with a defected smectic configuration at higher densities (not shown). 
By comparing the uniaxial and tetratic maps for density $\varphi=0.75$, Fig. \ref{fig1}(a),
we can see that the smectic domains in the experiments at high densities are located between neighbouring defects and 
close to the wall. As shown in our previous work, the presence of smectic textures at high packing fractions 
in granular rods with relative small aspect ratio 
is due to strong particle clustering promoted by a local energy dissipation mechanism \cite{us1}.

In order to understand the 
similarities and differences between the two systems in more detail, we have analysed the average defect behaviour. 
First we computed the position ${\bm r}=(x,y)$ 
of the point defects using the $q_4$ order parameter by identifying those particles 
with $q_4<0.2$; in strongly-developed tetratic configurations such as the ones shown in Fig. \ref{fig1}. This 
protocol leads to
four well-separated groups of particles associated with each of the four regions where the tetratic order is depleted. 
The centre of mass of each group of particles, ${\bm r}$, is taken as the position of the corresponding defect. The same procedure
is implemented in both experiments and MC simulation.

In the following the lab frame is assumed to be placed at the centre of the circular cavity, and polar coordinates 
${\bm r}=(r\cos{\theta},r\sin{\theta})$ for the position of the defects
are used to analyse various trends and distributions. The unwanted tendency of the system to
rotate \cite{Aranson3}, both in experiments and simulations, is suppressed by calculating the instantaneous average of the azimuthal angle over 
the four defects and subtracting this
angle from the angular position of each defect. This process isolates the inherent fluctuations of the defects about a mean position
by referring their
motion to a frame that rotates rigidly with the sample. Fig. \ref{fig2} shows the sampled positions of the defects in this frame
for both systems at packing fraction $\varphi=0.75$ (in the following results are presented
only for this case, since all the densities
explored, in the range $\varphi=0.70-0.75$, are qualitatively similar, while for $\varphi>0.75$  
large clusters in smectic-like configurations strongly compete with the tetratic ordering). 
Defects are colour-coded according to either the experimental time or the `MC time'. 
The plot shows that the system is being sampled ergodically. 
Sampling in the experimental system is comparatively poorer. 

The time evolution of defect positions in the experiment and in the MC simulation are shown in detail in
Fig. \ref{fig3}. Panels (a)-(c) correspond to the experiment, while panels (d)-(f) show the 
MC results. Fig. \ref{fig3}(a) shows the time evolution of the defect
angles $\theta_i$, $i=1,\cdots,4$, and also of the average
angle with respect to the lab frame. 
In this particular experiment the sample rotates globally in one direction,
but this is not the case in general.
Clearly defects slowly rotate with approximately constant angular
velocity $\omega$ and with superimposed fluctuations. Particle tracking and velocity calculations
indicate that this rotation is very rigid (i.e. linear velocity is roughly proportional to radial distance).
Some of these fluctuations are correlated, meaning that the solid-body rotation
occurs at nonconstant angular velocity $\omega(t)$. 

Fig. \ref{fig3}(b) shows the angles $\theta_i$
with respect to the frame with axes rotating with $\omega(t)$; in this frame individual defect
fluctuations about their mean are isolated and can be properly analysed. The radial distance of the defects also fluctuates in 
time, Fig. \ref{fig3}(c). In general defects stay close to the surfaces, which results from the repulsive defect interactions (we further comment
on this point below). 
MC simulation results, shown in Figs. \ref{fig3}(d)-(f), are qualitatively similar. Note that the poorer
experimental sampling due to practical time restrictions limit the temporal extent of fluctuations with 
respect to the simulation.

\begin{figure*}
\begin{center}
\includegraphics[width=0.32\linewidth,angle=0]{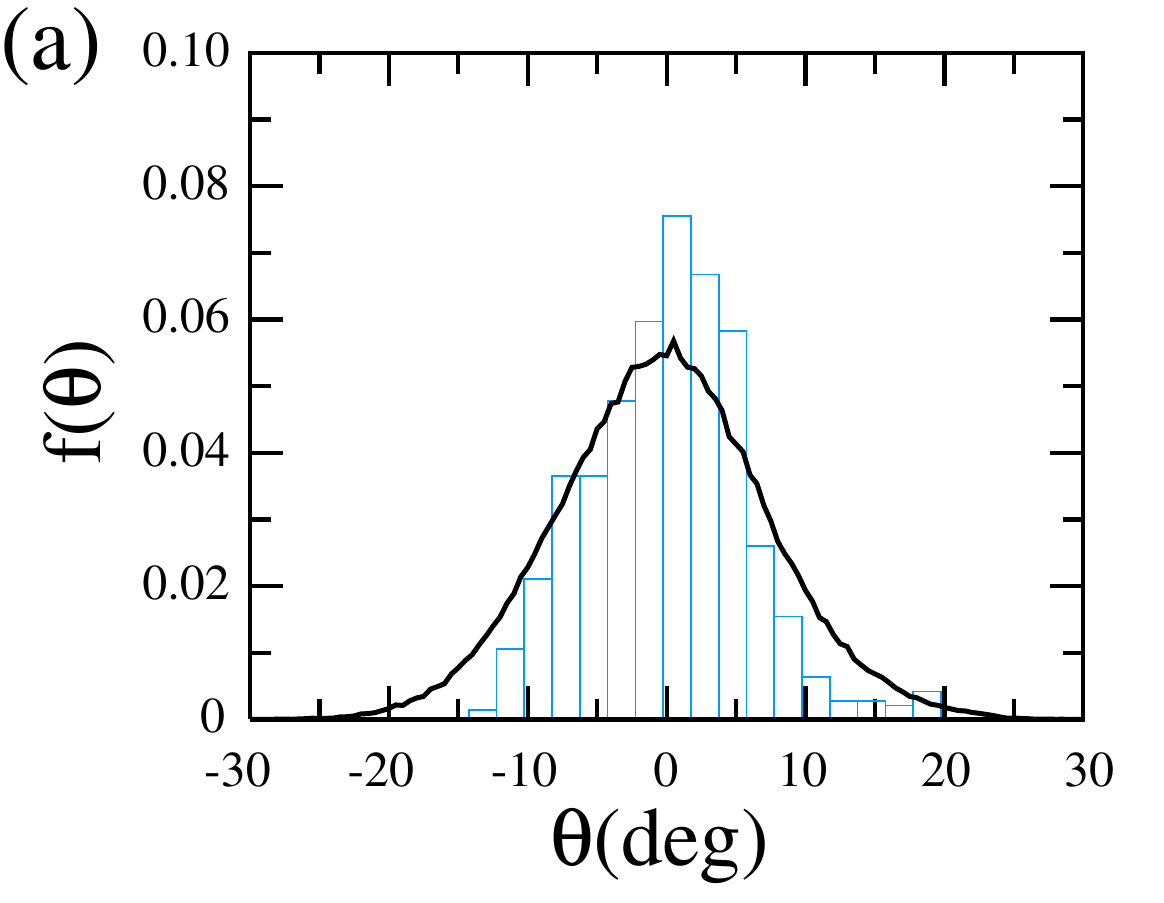}
\includegraphics[width=0.32\linewidth,angle=0]{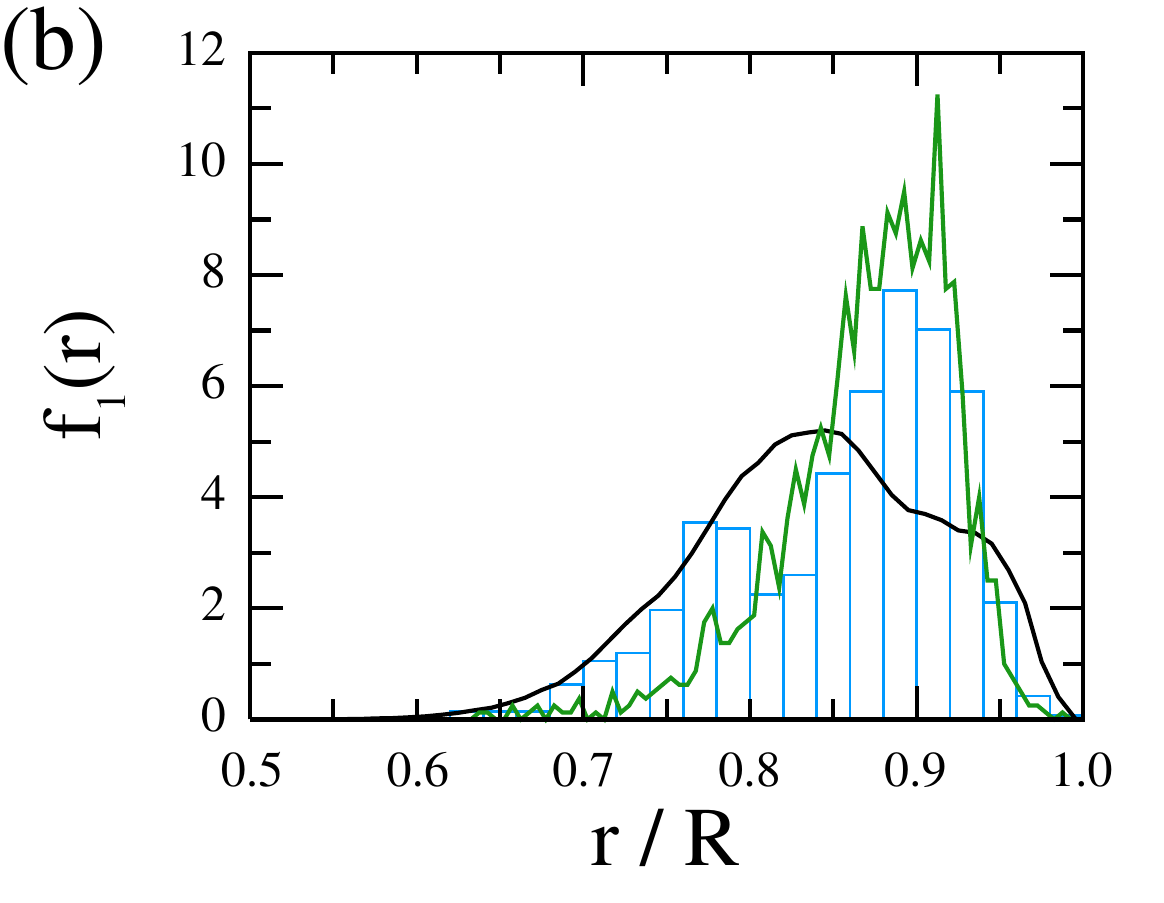}
\includegraphics[width=0.32\linewidth,angle=0]{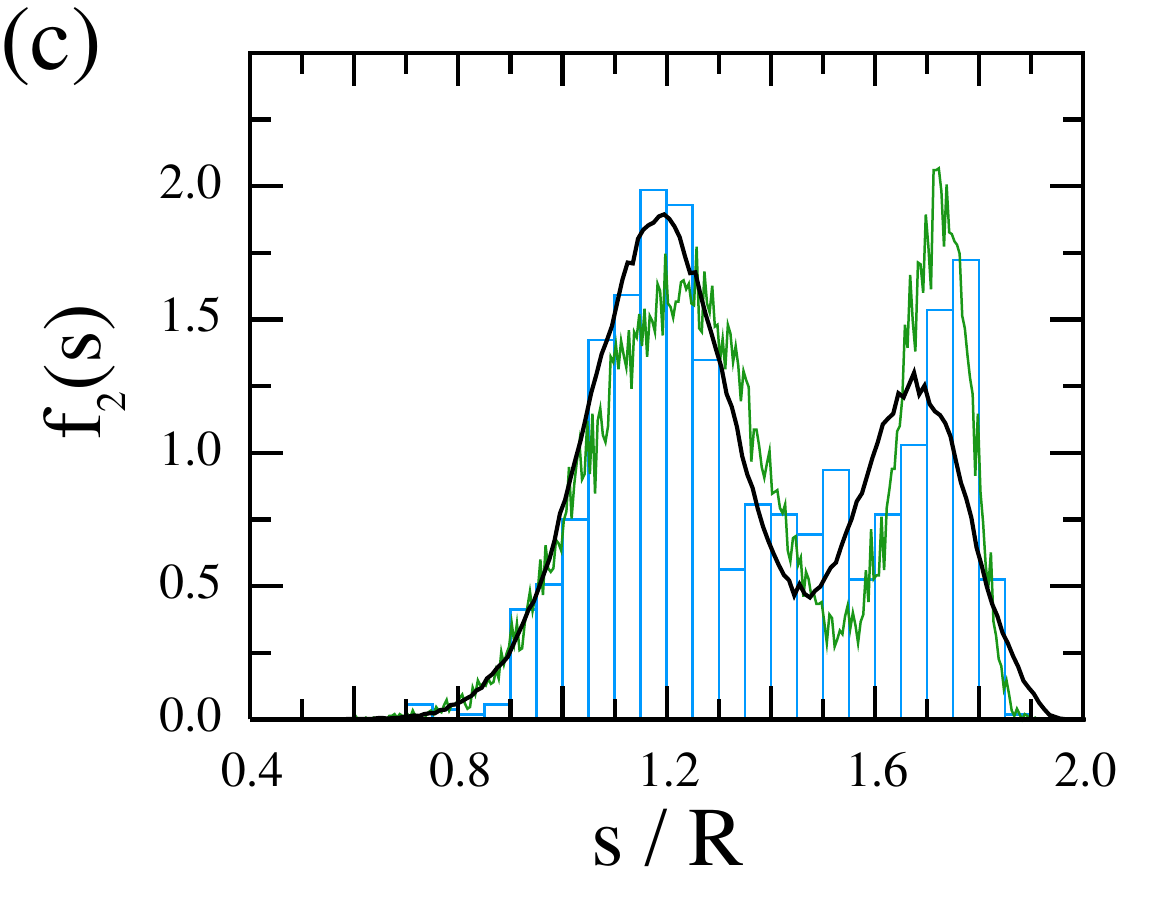}
\caption{\label{fig4}(a) Azimuthal angle distribution $f(\theta)$
from experiment (histogram) and simulation (black curve). 
(b) Radial distance distribution $f_1(r)$ from
experiment (histogram) and simulation (black curve), with respect to radial distance scaled with cavity radius. 
(c) Distribution of the relative distance between defects, $f_2(s)$, with
respect to the interdefect distance scaled with cavity radius.
Histogram: experimental results. Black curve: simulation. 
Green curves
in (b) and (c): Brownian simulation using the best-fit method 
applied on $f_2(s)$, as explained in the text.}
\end{center}
\end{figure*}

To show the fluctuation dynamics of the defects in more detail, 
the distribution of azimuthal angle, $f(\theta)$, is plotted in 
Fig. \ref{fig4}(a). Defect libration closely follows a Gaussian distribution in both systems, but the MC result 
presents a slightly broader distribution, although this is not conclusive
due to the relatively poorer statistics in the experiment.
The radial distance distributions $f_1(r)$
in the granular and thermal systems are also qualitatively similar, see Fig.~\ref{fig4}(b). They are not Gaussian but bimodal. In the case of the experiment 
this is clearly seen in Fig. \ref{fig3}(c), where the position of the two favoured distances is indicated by horizontal dotted lines.
Our interpretation for the bimodality is the following: even though defects repel each other, particles do form an
ordered surface structure that modifies the wall-defect interaction and prevents defects from reaching the wall. 
However, when this surface layer is absent, defects 
can be in close contact with the surface, resulting in the existence
of two favoured distances from the wall. The thickness of the surface layer is mainly determined by the density, but 
it differs substantially in the experiment and MC simulations, since the favoured particle surface orientation is different.
In the experiment, the orientation is mostly along the wall normal, with a thickness of one or two particle lengths. 
In the MC simulations, by contrast, the orientation is mostly planar \cite{Heras},
with a thickness of a few particle widths.
In both cases the average defect position along the radial distance is determined by the competition between the repulsive defect-defect 
interaction, which tends to push the defects to the wall, and the repulsive surface-defect interaction, which depends
on the particle orientation at the wall. The final distribution is bistable, but the location of the two maxima
is different in the experiment and in the MC simulation because of the different particle orientation favoured at the wall.

To better characterise the defect dynamics, we have also calculated the distribution of relative distances
between the defects, $f_2(s)$, which is shown in Fig. \ref{fig4}(c). As expected, it is also bimodal and exhibits two maxima, corresponding
to the nearest neighbour and next-nearest neighbour distances. The ratio between the two is very close to $\sqrt{2}$,
which confirms that the defects on average form a square
configuration. This arrangement is again the result of the repulsive surface-defect, and defect-defect interactions.
The distribution can be regarded as a superposition of two distributions located at the two characteristic distances.

The similarities between the MC simulations and the granular experiment shown above as regards the
distribution functions $f_i$ suggests that the nonequilibrium fluctuations of the order-parameter field in the 
experimental system can be assimilated to that of an effective system of Brownian point defects interacting 
through a fluctuating elastic medium. The medium would be characterised by means of an elastic constant. 
The mapping is natural for the equilibrium MC simulation. However, in the case of the experiment 
this mapping is not clear. The connection can be realised by relating
the fluctuations of the order-parameter field in the experiment, scaled
with an effective temperature, to that of the equilibrium system, measured through the usual
elastic constant scaled with the temperature.
Fluctuations in the experiment certainly depend on dissipation mechanisms such as inelastic particle 
interactions or friction between particles. Also, 
the definition of temperature in a granular system is controversial, and several choices can be found in the literature.
Therefore we suggest, as a working hypothesis, that the the strength of
elastic interactions, scaled with an effective temperature, in the granular and equilibrium
systems, are similar. We used this scaled elastic constant estimation as a way to quantify these interactions.
The mapping is a mere hypothesis, and the precise value of the scaled elastic constant
in the granular system might be different to that obtained below. Therefore, our result should be treated as a 
conjecture to be confirmed in the future by other direct methods.

We have implemented this procedure to extract information about the elastic stiffness coefficient of the tetratic
medium. In fact, the use of elastic-theoretical concepts in the context of granular monolayers is not new.
Galanis et al. \cite{Galanis2} applied elastic theory on the global orientation field of a vibrated monolayer
to infer the elastic behaviour of a monolayer consisting of very long rod in a uniaxial nematic configuration. However, 
defects were omitted from the analysis. Here we proceed differently, and assume the defect dynamics shown in the 
previous paragraphs can be modeled by equilibrium thermal fluctuations of an effective system of four
Brownian particles interacting through logarithmic potentials with no intervening medium. 
A numerical value for the conjectured elastic stiffness coefficient $K$ of the tetratic medium is then extracted 
by comparing the defect distribution of the experiment to that of the thermal effective model, explored by means of Brownian simulation. 

As pointed out before, the model has three free parameters: $T^*$, 
$\lambda^*=\lambda R$, and $\epsilon^*=\epsilon/c$.
To obtain their values
we focus on the relative-distance distribution $f_2(s)$ and define a best-fit function
in terms of the distance-integrated square difference between the experimental and the time-averaged
Brownian distribution extracted from simulation 
with fixed parameters $T^*,\lambda^*,\epsilon^*$. 
The values of these parameters are then
optimised by minimising this function (note that the simulation results do not depend on the scaled inverse friction
coefficient $\gamma$). The optimised values are $T^*\approx0.07$, $\lambda^*\approx 25$ and 
$\epsilon^*\approx 1.0$.

The optimised $f_2(s)$ function is shown in green in Fig. \ref{fig4}(c). 
The fitting is reasonable. Also in Fig. \ref{fig4}(b) the function $f_1(s)$
from the Brownian simulation is shown in green. Despite not being used
as a target function, the comparison with the experimental distribution
$f_1^{\rm(exp)}(r)$ is reasonable.

From the above optimised value of the parameters we obtain 
$c\simeq 14\;k_BT$. Since we expect the coefficient $c$ to be related to the elastic stiffness coefficient by 
$c=\pi k^2 K$ \cite{Bowick}, where $k=1/4$ is the winding number of
each of the four $+1$ defects, we obtain the elastic stiffness coefficient
as $K/k_BT\simeq 70$. We note that, 
to accommodate the fine structure of the experimental distribution 
$f_2^{\rm(exp)}(s)$, we also used a bistable 
wall potential $V(r)$. The resulting value for the scaled $K$ hardly changes, even though the fit certainly improves.

It is interesting to note that the value of the scaled 
$K$ obtained is remarkably close to the values of elastic coefficients
in two-dimensional nematic liquid-crystal phases. A two-dimensional nematic can only support splay and bend distortions, 
with elastic coefficients $K_1$ and $K_3$, respectively. In the tetratic phase symmetry imposes $K_1=K_3=K$. 
To our knowledge the only calculations of elastic constants in two-dimensional liquid crystals focused on rather
long rods (which can only exhibit uniaxial nematic ordering \cite{delasHeras}) at relatively low densities. 
The values have the same order of magnitude as the one obtained here.
Obviously a fitting of the Brownian model to the MC $f_2(s)$ function, which should be equivalent
to a direct calculation of $K$, gives a similar value.
However we note that, in the experiment, smectic fluctuations in the regions between 
neighbouring defects are frequent, as can be seen in Fig. \ref{fig1}. This adds an extra stiffness to the effective defect interaction,
a feature that could explain the differences observed in $f_2(s)$ between experiment and the equilibrium simulation.

In summary, we have shown that a vertically vibrated monolayer of granular rods can form configurations with tetratic
symmetries in a circular cavity. To restore the global symmetry broken by the cavity, the system develops four
localised defects close to the wall forming a square configuration. This is completely similar to the behaviour of the
equivalent thermal monolayer, which we have also investigated using MC simulation. In addition, the defect fluctuations 
about their average positions in the experimental and thermal systems are similar. We have exploited this
observation to conjecture that a properly scaled elastic stiffness coefficient for the granular monolayer 
should be similar to that of the equilibrium MC simulation. In order to exploit this idea, 
a Brownian dynamics simulation has been used to extract a value for such a scaled elastic constant.
The resulting value is close to those obtained from equilibrium theories on hard rods in two dimensions. Our 
results give evidence that vibrated monolayers of dissipative particles, at least in some window of experimental
conditions, have similar ordering properties, respond equally to symmetry-breaking external fields,  
and would possess elastic stiffness constants as in the corresponding thermal systems.

The fact that the number of defects and their spatial distributions depend on the 
symmetry of the 
ordered phases (nematic, tetratic or smectic) and on the geometrical restrictions imposed by confinement, makes
granular rod monolayers an ideal tool to device well controlled experiments with the aim to study the dependence of these properties on the system boundary conditions.
For example the design of a ring-shaped container could change the nature of 
stationary textures present in the system and also
the distribution of defects~\cite{Mulder}. 
Note that, in experiments on colloidal or molecular systems,
the presence of non-controlled heterogeneities at particle length scales
in the confining surfaces gives rise to important fluctuations that distort the orientational director field
and consequently the final distribution of defects.
Moreover, the study of liquid-crystalline textures and defects 
in granular systems is not limited to quasi-two-dimensional systems~\cite{Maza}.

\acknowledgments

Financial support under grants FIS2015-66523-P and FIS2017-86007-C3-1-P 
from Ministerio de
Econom\'{\i}a, Industria y Competitividad (MINECO) of Spain is acknowledged.


\begin{references}
\bibitem{Narayan} V. Narayan, N. Menon, and S. Ramaswamy, J. Stat. Mech. - Theory and Experiment, P01005 (2006). 
\bibitem{Galanis1} J. Galanis, D. Harries, D. L. Sackett, W. Losert, and R. Nossal, Phys. Rev. Lett. {\bf 96}, 028002 (2006).
\bibitem{Galanis2} J. Galanis, R. Nossal, W. Losert, and D. Harries, Phys. Rev. Lett. {\bf 105}, 168001 (2010).
\bibitem{Aranson1} I. S. Aranson, and L. S. Tsimring, Rev. Mod. Phys. {\bf 78}, 641 (2006).
\bibitem{Aranson2} E. Khain, and I. S. Aranson, Phys. Rev. E {\bf 84}, 031308 (2011).
\bibitem{Narayan2} V. Narayan, S. Ramaswamy, and N. Menon, Science {\bf 317}, 105 (2007).
\bibitem{Mueller} T. Muller, D. de las Heras, I. Rehberg, and K. Huang, Phys. Rev. E {\bf 91}, 062207 (2015).
\bibitem{us1} M. Gonz\'alez-Pinto, F. Borondo, Y. Mart\'{\i}nez-Rat\'on, and E. Velasco, Soft Matter {\bf 13}, 2571 (2017).
\bibitem{Schlacken} H. Schlacken, H.-J. Mogel, and P. Schiller, Mol. Phys. {\bf 93}, 777 (1998).
\bibitem{us0} Y. Mart\'{\i}nez-Rat\'on, E. Velasco, and L. Mederos, J. Chem. Phys. {\bf 122}, 064903 (2005).
\bibitem{Chaikin} K. Zhao, C. Harrison, D. Huse, W. B. Russel, and P. M. Chaikin, Phys. Rev. E {\bf 76}, 040401(R) (2007).
\bibitem{Donev} A. Donev, J. Burton, F. H. Stillinger, and S. Torquato, Phys. Rev. B {\bf 73}, 054109 (2006).
\bibitem{us2} Y. Mart\'{\i}nez-Rat\'on and E. Velasco,  Phys. Rev. E {\bf 79}, 011711 (2009).
\bibitem{Edwards} R. Blumenfeld, and S. F. Edwards, Eur. Phys. J. {\bf 223}, 2189 (2014).

\bibitem{LiBowick} These defects have also been theoretically studied by MC simulations and elastic free-energy calculations
on thermal systems with tetratic ordering on spherical surfaces, resulting in the presence of eight disclinations located
at the vertexes of an anticube. See Y. Li, H. Miao. H. Ma, and J. Z. Y. Chen, Soft Matter {\bf 9}, 11461 (2013), and
O. V. Manuhina and M. J. Bowick, Phys. Rev. Lett. {\bf 114}, 117801 (2015).
\bibitem{us} Y. Mart\'{\i}nez-Rat\'on, E. Velasco, and L. Mederos, J. Chem. Phys. {\bf 125}, 014501 (2006).

\bibitem{ImageJ} {\tt https://imagej.nih.gov/ij/}
\bibitem{Heras} D. de las Heras and E. Velasco, Soft Matter {\bf 10}, 1758 (2014).
\bibitem{Bowick} M. J. Bowick and L. Giomi, Adv. Phys. {\bf 58}, 449 (2009).
\bibitem{Aranson3} I. S. Aranson, D. Volfson, and L. S. Tsimring, Phys. Rev. E {\bf 75}, 051301 (2007).
\bibitem{delasHeras} D. de las Heras, L. Mederos, and E. Velasco, Liq. Crys. {\bf 37}, 45 (2009).



\bibitem{Mulder} I.C. G{\^a}rlea, P. Mulder, J. Alvarado, O. Dammone, D.G.A.L. Aarts, M.P. Lettinga, G.H. Koenderink, and B.M. Mulder, Nat. Commun. {\bf 7}, 12112 (2016).
\bibitem{Maza} K. Asencio, M. Acevedo, I. Zuriguel, D. Maza, Phys. Rev. Lett. {\bf 119}, 228002 (2017).
\end{references}
\end{document}